\documentclass[twocolumn]{aastex701}
\usepackage{graphicx}
\usepackage{overpic}
\usepackage{multirow}
\usepackage{array}
\usepackage{subcaption}
\usepackage{xcolor}
\usepackage{longtable}
\usepackage{newtxtext}
\usepackage{newtxmath}
\usepackage{hyperref}
\usepackage[version=4]{mhchem}
\usepackage{float}
\usepackage{placeins}

\begin{document}
\title{JWST Edge-on Disk Ice (JEDIce): Vibrationally hot, rotationally cold H$_2$ in the outer disk of Oph 163131 non-thermally excited by
UV and cosmic rays}
\author[orcid=0000-0002-2131-4346]{Korash Assani}
\affiliation{Department of Astronomy, University of Virginia, Charlottesville, VA 22903, USA}
\affiliation{Virginia Institute of Theoretical Astronomy, University of Virginia, Charlottesville, VA 22903, USA}
\email{ka8km@virginia.edu}

\author[0000-0002-7402-6487]{Zhi-Yun Li}
\affiliation{Department of Astronomy, University of Virginia, Charlottesville, VA 22903, USA}
\affiliation{Virginia Institute of Theoretical Astronomy, University of Virginia, Charlottesville, VA 22903, USA}
\email{zl4h@virginia.edu}

\author[0000-0002-8716-0482]{Jennifer B.~Bergner}
\affiliation{University of California, Berkeley, Berkeley CA 94720, USA}
\email{jbergner@berkeley.edu}

\author[orcid=0000-0001-8341-1646]{David A. Neufeld}
\affiliation{Department of Physics \& Astronomy, Johns Hopkins University, Baltimore, MD 21218, USA}
\email{neufeld@jhu.edu}

\author[0000-0001-6307-4195]{Daniel Harsono}
\affiliation{Institute of Astronomy, Department of Physics, National Tsing Hua University, Hsinchu, Taiwan}
\email{dharsono@gapp.nthu.edu.tw}

\author[0000-0001-7479-4948]{Maria N. Drozdovskaya}
\affiliation{Physikalisch-Meteorologisches Observatorium Davos und Weltstrahlungszentrum (PMOD/WRC), Dorfstrasse 33, 7260 Davos Dorf,
Switzerland}
\affiliation{Department of Chemistry, Biochemistry and Pharmaceutical Sciences (DCBP), Universit{\"a}t Bern
Freiestrasse 3, 3012 Bern, Switzerland}
\email{maria.drozdovskaya.space@gmail.com}

\author[orcid=0000-0003-2303-0096]{Marco Padovani}
\affiliation{INAF-Osservatorio Astrofisico di Arcetri, Largo E. Fermi 5, 50125 Firenze, Italy}
\email{marco.padovani@inaf.it}

\author[orcid=0000-0003-1197-7143]{Emmanuel Dartois}
\affiliation{Institut des Sciences Mol\'eculaires d'Orsay (ISMO), CNRS \& Universit\'e Paris$-$Saclay, Orsay, France}
\email{emmanuel.dartois@universite-paris-saclay.fr}

\author[0000-0003-4985-8254]{Jennifer A. Noble}
\affil{Physique des Interactions Ioniques et Mol\'{e}culaires, CNRS, Aix-Marseille Univ., 13397 Marseille, France}
\email{jennifer.noble@univ-amu.fr}

\author[0000-0003-2631-5265]{Nicole Arulanantham}
\affiliation{Astrophysics \& Space Institute, Schmidt Sciences, New York, NY 10011, USA}
\email{narulanantham@schmidtsciences.org}

\author[0000-0003-2014-2121]{Alice S. Booth} 
\affiliation{Center for Astrophysics \textbar\, Harvard \& Smithsonian, 60 Garden St., Cambridge, MA 02138, USA}
\email{alice.booth@cfa.harvard.edu}

\author[0000-0001-8227-2816]{Yao-Lun Yang}
\affiliation{Star and Planet Formation Laboratory, RIKEN Pioneering Research Institute, Wako-shi, Saitama, 351-0106, Japan}
\email{yaolunyang.astro@gmail.com}

\author[0000-0002-0554-1151]{Mayank Narang}
\affiliation{Jet Propulsion Laboratory, California Institute of Technology, 4800 Oak Grove Drive, Pasadena, CA 91109, USA}
\email{mayank.narang@jpl.nasa.gov}

\author[0000-0002-0786-7307]{Will E. Thompson}
\affiliation{University of California, Berkeley, Berkeley CA 94720, USA}
\email{willet@berkeley.edu}

\author[0000-0002-3401-5660]{Elizabeth Yunerman}
\affiliation{Center for Astrophysics, Harvard \& Smithsonian, 60 Garden St., Cambridge, MA 02138, USA}
\email{eyunerman@cfa.harvard.edu}

\author[0000-0001-8798-1347]{Karin I. \"Oberg}
\affiliation{Center for Astrophysics, Harvard \& Smithsonian, 60 Garden St., Cambridge, MA 02138, USA}
\email{koberg@cfa.harvard.edu}

\author[0000-0002-3401-5660]{Julia C.~Santos}
\affiliation{Center for Astrophysics, Harvard \& Smithsonian, 60 Garden St., Cambridge, MA 02138, USA}
\email{julia.santos@cfa.harvard.edu}

\author[0000-0002-8023-2834]{Charles Mentzer}
\affiliation{Institute of Astronomy, Department of Physics, National Tsing Hua University, Hsinchu, Taiwan}
\email{cmentzer@gapp.nthu.edu.tw}

\author[orcid=0000-0002-3835-3990]{Jon P. Ramsey}
\affiliation{Bluedrop Training \& Simulation, Inc., 36 Solutions Drive \#300, Halifax, Nova Scotia B3S 1N2, Canada }
\email{}

\author[0009-0008-8810-6577]{Lukas Welzel}
\affiliation{Leiden Observatory, Leiden University, PO Box 9513, 2300 RA Leiden, The Netherlands}
\email{welzel@strw.leidenuniv.nl}

\author[0000-0001-7552-1562]{Klaus M. Pontoppidan}
\affiliation{Jet Propulsion Laboratory, California Institute of Technology, 4800 Oak Grove Drive, Pasadena, CA 91109, USA}
\email{klaus.m.pontoppidan@jpl.nasa.gov}

\author[0000-0003-1878-327X]{Melissa McClure}
\affiliation{Leiden Observatory, Leiden University, PO Box 9513, 2300 RA Leiden, The Netherlands}
\email{mcclure@strw.leidenuniv.nl}

\begin{abstract}
Constraining ionization and excitation processes in protoplanetary disks is essential for understanding the chemical structure and evolution of disk material, shaping planet formation pathways. We present JWST/NIRSpec IFU observations of the edge-on disk Oph 163131, which reveal a unusual ro-vibrational H$_2$ spectrum dominated by the 1--0 O(2) line (2.627 $\mu$m), with suppressed higher-$J$ emission despite excitation to $v=2$ and $3$. This vibrationally hot, rotationally cold H$_2$ emission is spatially extended, broadly following the molecular disk traced by CO($J{=}2$--1), with emission increasing above and below a thin midplane dark lane and extending radially beyond $\sim$200 au, where near-IR scattered-light emission is no longer dominant. We interpret the observed H$_2$ emission as arising from non-thermal excitation in cold, dense outer-disk gas, where collisions depopulate higher-$J$ rotational levels within each vibrational manifold prior to emission, producing the characteristic ``$v$-hot, $J$-cold" spectrum. We consider both ultraviolet irradiation and cosmic-ray excitation as contributors to the H$_2$ emission and find that their combined action, together with collisional de-excitation of high-$J$ level populations, broadly reproduces the observed line ratios and morphology. Within this framework, we infer a rather high effective cosmic-ray ionization rate of $\sim(1$–$10)\times10^{-15}$ s$^{-1}$ in the presence of a moderate UV field ($\chi_{UV}=100-1000$, in Draine units). These results for disks, together with the recent findings by Bialy et al. 2025 for the lower-density starless core B68, highlight the potential of ro-vibrational H$_2$ emission as a novel probe of cosmic-ray ionization.

\end{abstract}

\received{February 02, 2026}
\accepted{July 08, 2026}
\section{Introduction}\label{sec:introduction}


The irradiation and ionization environments of protoplanetary disks play a central role in influencing their structure and the chemical and physical evolution of disk material from which planets form \citep[e.g.][]{Alexander2014_PPVI,Gorti_photoevaporative_wind_FUV_EUV_Xray_2007,ErcolanoPascucci2017_Review}. However, the relative importance and spatial extent of these processes throughout the disk remain poorly constrained observationally; for example, it is often unclear which disk regions are primarily influenced by local sources associated with the central protostar and shocks versus external influences such as nearby massive stars and galactic cosmic rays. With JWST's impressive sensitivity to very dim spectral lines and its sub-arcsecond integral-field spectroscopy, we can identify and spatially analyze numerous infrared tracers that probe the excitation environment down to tens of au in nearby star-forming regions (within a few hundred parsecs). This allows us to determine where different species emit or absorb light and the strength at which they do so across multiple transitions, and then use physical and chemical models to constrain the processes responsible for the observed spectra. 

In edge-on protoplanetary disks, the optically thick disk midplane obscures most or all direct stellar light from the observer, providing a unique opportunity to disentangle the spatial origin of disk emission. At shorter wavelengths (i.e., optical and infrared), the observed continuum emission can be dominated by starlight scattered by small dust grains in the disk surface layers, often producing a bipolar reflection nebula separated by a dark midplane in edge-on systems \citep[e.g.][]{Dartois_flyingsaucer_jedice_2025, Duchene_EODs_mixed_grains_outer_disk_2024, Takazi_JWST_edge_on_imaging_mid_IR_dust_scattering_2025, Bergner_JEDIce_program_oveview_2026}. In contrast, longer-wavelength (i.e., sub-mm/mm) continuum emission primarily traces thermal emission from a more vertically settled dust population concentrated toward the disk midplane, potentially including larger grains \citep[e.g.][]{Villenave_EOD_alma_B47_2019, Villenave_highly_setled_Oph163131_2022}. This viewing geometry turns out to be very useful for distinguishing between the scattered-light disk surface traced by the continuum emission at a given wavelength and spatially extended gas emission beyond the disk surface, such as emission associated with outflowing material \citep[e.g.][]{pascucci2024nested}. Here, we use this geometry to investigate ro-vibrational H$_2$ emission from the outer gas disk of an edge-on protoplanetary disk and explore, for the first time, its use as a direct probe of both ultraviolet irradiation and cosmic-ray ionization within the disk.

Ro-vibrational H$_2$ emission can be produced by ultraviolet (UV) radiative pumping, thermal excitation in warm gas (e.g., shocks), or non-thermal excitation mediated by energetic particles such as secondary electrons generated by cosmic-ray (CR) ionization \citep[e.g.][]{Black_vanDishoeck_1987, Gredel_infrared_H2_Xrays_1995, nomura05, nomura07}. A key aspect common to all of these non-thermal excitation pathways is the conservation of the ortho–para nuclear spin symmetry of H$_2$, such that para lines originate from excitation of molecules primarily in the $v=0$, $J=0$ ground state, while ortho lines originate from excitation of molecules initially in the $v=0$, $J=1$ state ($E_{v=0,J=1}/k\sim$171 K, where k is Boltzmann's constant), making their relative strengths sensitive to the thermal history of the gas (i.e., ortho lines preferentially trace warm or previously heated gas).

In strongly irradiated star-forming environments such as photodissociation regions (PDRs), near-infrared H$_2$ emission is expected to arise from UV pumping through the Lyman–Werner bands, in which UV photons, both, heat the gas and excite H$_2$, predominantly from the lowest rotational states ($v=0$, $J=0$ and $J=1$), to highly excited vibrational levels ($v\ge1$) that then radiatively cascade \citep[e.g.][]{Black_vanDishoeck_1987}. This produces emission spanning both ortho- and para-rotational levels and across multiple vibrational manifolds. In PDRs, such as those observed with JWST toward the Orion Bar and Horsehead Nebula, ro-vibrational H$_2$ emission peaks at the H/H$_2$ dissociation front, where molecular hydrogen survives UV irradiation due to attenuation and self-shielding, while remaining efficiently excited \citep{Peeters_PDRs4All_orionbar_2024, Zannese_PDRs4all_horsehead_H2_2025, Zannese_PDRs4All_HD_Orion_bar_2026}.

In contrast, secondary electrons, produced by the ionization of molecular hydrogen induced by cosmic rays (CRs), collisionally excite H$_2$, influencing the resulting ro-vibrational emission. Unlike UV photons, CRs can penetrate to larger column densities \citep{ Padovani_CR_molec_clouds_2009, draine2010physics}, potentially exciting H$_2$ in cold, well-shielded gas that is predominantly in the $v{=}0$, $J{=}0$ para (even) state. Under these conditions, CR-driven excitation is expected to preferentially enhance even-$J$ (para-H$_2$) transitions, predicting relatively bright lines such as v=1–0 O(2) compared to commonly observed ortho-H$_2$ lines such as S(1) and O(3). This behavior has been explored in theoretical studies assessing whether JWST-accessible near-infrared ro-vibrational H$_2$ line ratios can serve as diagnostics of CR-driven excitation under favorable conditions \citep[e.g.][]{Bialy_cold_clouds_CR_detecttors_2020, Padovani_CR_molec_clouds_2009,Padovani_lowenergyCRs_2020,Padovani_H2_CR_with_JWST_models_2022, Gaches_CR_H2_models_2022, Bialy_CR_NIR_dense_clouds_testbed_2022}. 

Observational support for CR-driven excitation of ro-vibrational H$_2$ has recently been found toward the starless core Barnard 68 (B68), where JWST observations \citep{Bialy2025_B68_CRH2,Neufeld2025_CRinB68_JWST} revealed enhanced para-H$_2$ emission consistent with theoretical expectations. \citet{Bialy2025_B68_CRH2} inferred a CR ionization rate of order $\zeta \sim 10^{-16}$  $\mathrm{s^{-1}}$, which is elevated relative to values typically inferred for dense molecular gas in molecular clouds and the Galactic ISM ($\zeta \sim \text{few} \times 10^{-17}$  $\mathrm{s^{-1}}$; \citealt{Dalgarno_interstellar_galactic_CR_rates_2006,Padovani_CR_molec_clouds_2009,Indriolo_CR_rate_galactic_diffuse_ISM_2012}). 

While B68 represents a cold, starless environment distinct from protoplanetary disks, these results demonstrate that near-infrared H$_2$ diagnostics are sensitive to enhanced CR excitation under favorable conditions. In protoplanetary disks, however, the effective CR ionization rate may be strongly reduced by magnetized winds or outflows, potentially limiting the applicability of these diagnostics \citep[e.g.][]{cleeves_xray_v_CR_rates_TW_Hya_2015}. Whether similar CR-sensitive H$_2$ diagnostics can be applied to protoplanetary disks, where gas densities are higher and post-excitation collisional redistribution of level populations is more efficient, remains an open question that this study directly addresses.

In this paper, we use JWST/NIRSpec IFU observations of the edge-on disk Oph 163131 to explore the excitation regimes reponsible for ro-vibrational H$_2$ emission. Oph 163131 (SSTc2d J163131.2–242627) is one of the most highly settled disks with millimeter dust emission confined to a geometrically thin midplane layer with a scale height of $\lesssim$0.5 au at 100 au \citep{Villenave_highly_setled_Oph163131_2022}. In contrast, $^{12}$CO observations trace a more extended gaseous disk with elevated brightness temperatures ($T\sim30$–40 K at radii beyond 200 au from the host protostar), consistent with cold outer-disk gas exposed to external ultraviolet irradiation \citep{Flores_unusualanatomy_oph163131_2021}. This system, therefore, provides a well-controlled disk environment for assessing how irradiation, ionization, and collisional redistribution shape near-infrared H$_2$ excitation. In Section \ref{sec:observations_data_Reduction}, we describe the observations and data reduction. In Section \ref{sec:3_rovibexcitation}, we evaluate ultraviolet and cosmic-ray excitation scenarios and assess how post-excitation collisional de-excitation reshapes the observed H$_2$ line ratios. Section \ref{sec:discussion} places these results in a broader physical context, before concluding in Section \ref{sec:conclusion}.

\section{Observation Results}
\label{sec:observations_data_Reduction}

We present JWST/NIRSpec IFU observations of Oph 163131 obtained with the G235H and G395H gratings, covering the 1.7--4 $\mu$m region, where strong near-IR ro-vibrational H$_2$ lines ($v$=1--0, 2--1, 3--2 in S, Q, and O branches) reside. In the high-resolution grating settings, however, the brightest Q-branch transitions (2.40--2.43~$\mu$m) unfortunately fall within the detector gap. The observations were taken as part of the Cycle 3 JWST program \textit{JWST Edge-on Disk Ice} (JEDIce, PID 5299, PI Bergner), which targets a sample of edge-on disks and embedded young stellar objects \citep{Bergner_JEDIce_program_oveview_2026}. The NIRSpec IFU provides a $3\arcsec\times3\arcsec$ field of view with spatial sampling of $0\farcs1$, corresponding to $\sim15$ au at the distance of $\rho$ Oph \citep[$\sim147 \rm pc$,][]{Ortiz_leon_gould_belt_distances_oph_2017, Ortiz_Leon_GaiaDR2_oph_serpens_aquila_2018}, and spectral resolving power $R\sim2700$ in the wavelength range of interest.

The data were processed with the JWST Calibration Pipeline v1.17.1 \citep{Bushouse_1_17_1_jwst_pipeline} and following the post-reduction calibration steps that are discussed in detail in \citep{Bergner_JEDIce_program_oveview_2026}. We performed additional post-processing to subtract continuum emission and isolate line surface brightness. For each transition of interest, we fit and removed a local linear baseline and integrated the residual over a fixed-width frequency window to construct moment-0 maps. For the spectral analysis, we extracted a spectrum in the bright outer-disk region, where the 1--0 O(2) emission is strong and coincides with the CO gas contours (see \ref{subsec:spectral}). The IFU maps were aligned with the ALMA Band~6 continuum and $^{12}$CO $J=2-1$ emission (230.54 GHz) using the target astrometry and continuum cross-correlation. The Band~6 continuum image was obtained from the ALMA archive from the 2018 Cycle 6 observations (2018.1.00958.S; PI: Villenave), with a beam size of $\sim$$0.02^{\arcsec}$. The comparison between CO and H$_2$ in the following sections uses the final continuum-subtracted $^{12}$CO $J=2-1$ cube from \citet{Villenave_highly_setled_Oph163131_2022}, which combines the 2016 Cycle 4 (2016.1.00771.S; PI: Duchêne) and 2018 Cycle 6 observations and has a synthesized beam of $0.081^{\arcsec}\times0.072^{\arcsec}$. This is comparable to the JWST NIRSpec IFU angular resolution of $\sim0.1^{\arcsec}$, corresponding to $\sim15$ au at 147 pc.

\subsection{Morphology of H$_2$ Emission}
\label{subsec:morphology}

\begin{figure*}[t]
    \centering
    \includegraphics[width=0.9\linewidth, trim={0cm 0cm 0.0cm 0cm},clip]{ 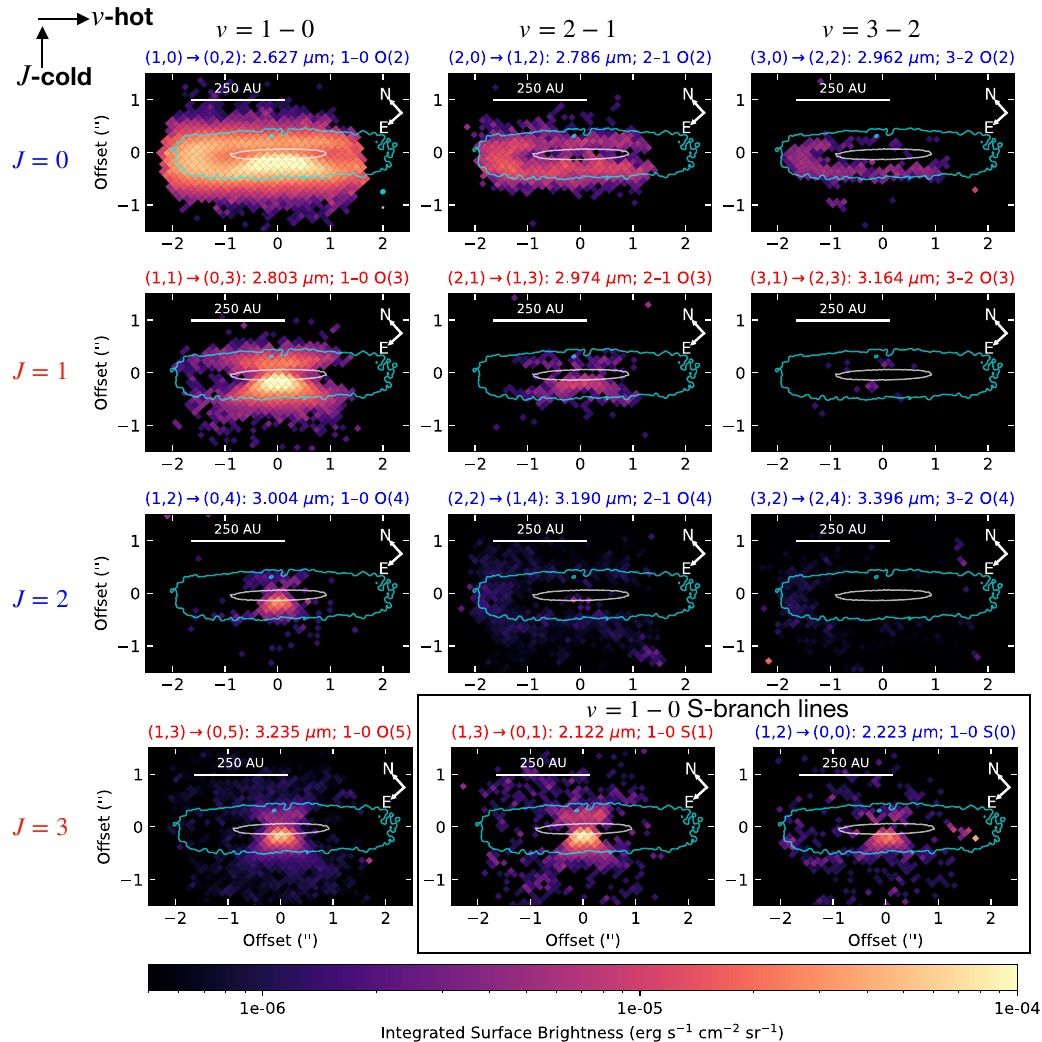}

\caption{
Integrated moment--0 maps of multiple H$_2$ ro-vibrational emission lines toward Oph 163131, rotated by $40^{\circ}$ to align the disk midplane horizontally.  Panel titles list the $(v_u,J_u)\rightarrow(v_l,J_l)$ transitions and wavelengths. Columns correspond primarily to the $v=1$--0, $v=2$--1, and $v=3$--2 bands (left to right), and rows are ordered by increasing upper rotational quantum number $J_u=0$--3 (top to bottom); the $v=1$--0 S(0) and S(1) lines are shown in the final row for comparison with other $v=1$--0 O-branch transitions. White contours trace the ALMA Band~6 dust continuum, and cyan contours show CO $J=2$--1 emission, with beam sizes indicated in the top-left panel towards the bottom right. The first row highlights the significant emission from highly excited vibrational manifold $v=2$ and $3$ (i.e., $v$-hot), particularly near the left (NE) end of the disk, while the first column showcases the progressively weaker emission from higher-$J$ levels within the $v=1$ manifold (i.e., $J$-cold).}
\label{fig:fig1}

\end{figure*}

Figure \ref{fig:fig1} shows integrated moment--0 maps of several key H$_2$ ro-vibrational lines observed towards Oph 163131, overlaid with the ALMA Band 6 dust continuum (cyan contours) and CO $J{=}2$--1 emission (white contours) observations. The para 1--0 O(2) line at 2.627 $\mu$m (top left panel) is the brightest ro-vibrational transition. Its emission approaches the edge of the NIRSpec IFU observational FOV, suggesting that it likely extends even further. The MIRI 7.7 $\mu$m emission reported by \citet{Villenave_JWST_mid_IR_oph163131_2024} is clearly extended, even along the midplane, consistent with widespread surface-layer emission at larger projected radii. The MIRI IFU observations provide an independent test for extended warm H$_2$ emission at projected distances $\gtrsim 2.5''$ along the disk major axis (Assani et al. in prep.).

The outer-disk morphology is also evident in the higher-vibrational O(2) lines shown in the first row of Figure \ref{fig:fig1} (2–1 O(2) in the second column and 3–2 O(2) in the third), which more clearly show the O(2) surface brightness declining toward the edge of the FOV and strengthening toward the northeast (NE) side of the disk. These O(2) transitions are the three brightest lines that dominate the ro-vibrational spectrum, while other detected ro-vibrational lines originating from higher rotational levels (second through fourth rows of Figure \ref{fig:fig1}) are significantly fainter, particularly in the outer disk (r $\gtrsim$ 100 au away from the conical polar regions traced by scattered light). Higher-$J$ transitions such as v=1–0 O(5), S(1), and S(0) are detected more strongly within the scattered-light region and exhibit a more conical morphology. As a check, we also construct continuum maps by integrating the baseline fits over the same spectral window as the line emission and then produce line-to-continuum ratio maps (Figures \ref{fig:fig_app1} and \ref{fig:fig_app2} in the Appendix \ref{App:continuum_line_to_continuum_maps}). These show that although H$_2$ lines are detected within the scattered-light regions, the continuum emission there is relatively strong ($\gtrsim 10\times$ that of the line), making it more difficult to exclude the possibility that some fraction of the observed line emission is scattered into the aperture from regions closer to the star. These maps further confirm that the O(2) emission is spatially extended and that the line-to-continuum ratio becomes largest outside the scattered-light dominated part of the disk, supporting a more direct association between the detected H$_2$ emission and locally emitting outer-disk gas.

The extended O(2) emission broadly follows the CO($J=$2–1) gas disk shown as cyan contours in Figure \ref{fig:fig1}. The CO contours are drawn at 5$\sigma$ (or $\sim$4 mJy) to highlight the diffuse outer gas \citep[similar to][]{Villenave_highly_setled_Oph163131_2022}, though here we use a peak-intensity map rather than an integrated CO map. The H$_2$ emission is significantly weaker in the midplane region within a radius of $\sim$200 au, forming a dark lane most noticeable in the O(2) H$_2$ maps (top row of Figure \ref{fig:fig1}). The Band 6 dust continuum lane is also shown (white contours, Figure \ref{fig:fig1}), which is more compact than the CO-traced region, both, radially and vertically, indicating that the bulk of the dust responsible for the millimeter continuum is more concentrated than the gas, consistent with the extreme dust settling inferred by \citet{Villenave_highly_setled_Oph163131_2022}. It lies within the dark lane where the O(2) emission is significantly reduced relative to the thick bright sheath enveloping it. 

\subsection{Spectral Characteristics}\label{subsec:spectral}

\begin{figure*}[t]
    \centering
    \includegraphics[width=0.98\linewidth, trim={0.0cm 0cm 0.0cm 0cm},clip]{ 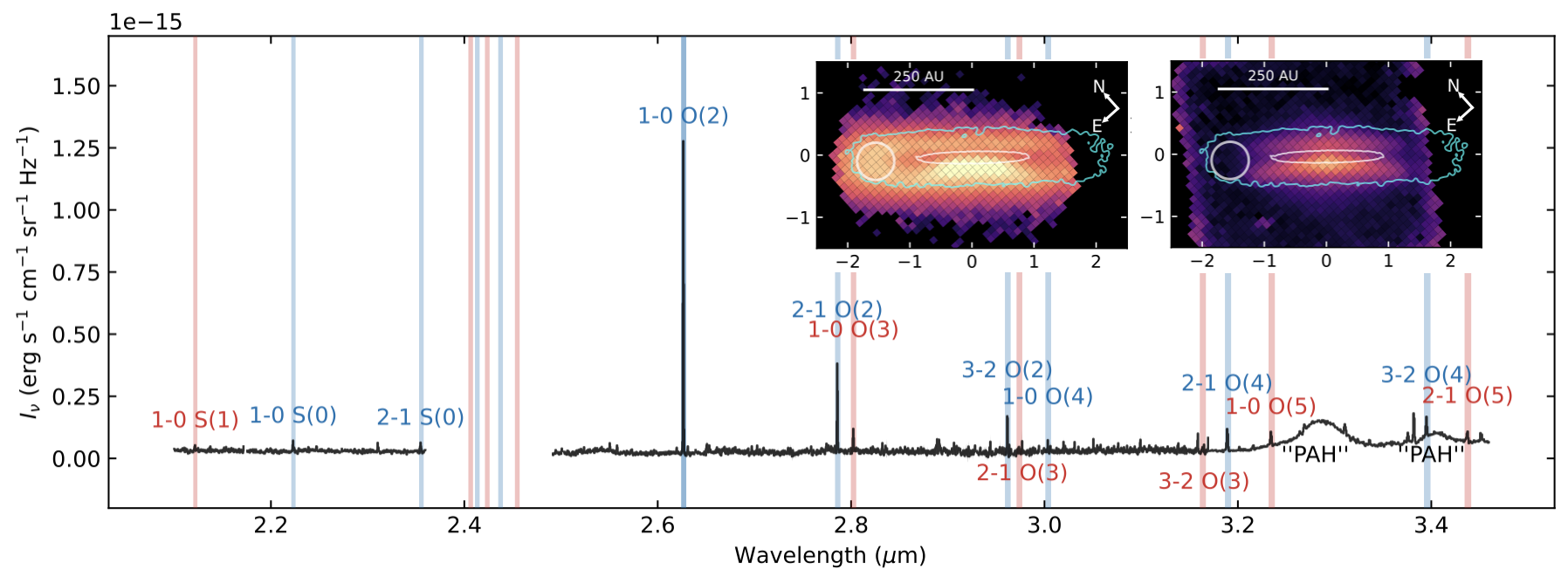}
\caption{
JWST near-infrared spectrum of Oph~163131 extracted within a circular aperture ($D=0.5''$). The spectrum shows the mean surface brightness averaged over all pixels inside the aperture. Vertical shaded bands mark the wavelengths of detected H$_2$ ro-vibrational transitions, with para-H$_2$ lines shown in blue and ortho-H$_2$ lines in orange; corresponding line labels are indicated above the spectrum. The $\sim$3.3--3.4 $\mu$m PAH emission features are also labeled. Inset panels show, for reference, the integrated surface-brightness map of the H$_2$ $v{=}1$--0
O(2) line at 2.627 $\mu$m (top left) and the corresponding continuum map (top right), with the extraction aperture overlaid.
}
\label{fig:fig2}
\end{figure*}
A representative spectrum from the outer disk is shown in Figure \ref{fig:fig2}, with the corresponding extraction aperture marked on the integrated H$_2$ and continuum maps. The 1--0 O(2) line at 2.627 $\mu$m is the dominant feature in the spectrum (upper level from $v=1$, $J=0$). The 1--0 O(3) line (from $v=1$, $J=1$) is clearly detected and is brighter than the very weak 1--0 S(1) line at 2.122 $\mu$m (from $v=1$, $J=3$). Transitions whose upper levels originate from $J=2$ (1--0 O(4), 1--0 S(0)) are also detected but remain faint, while the higher vibrational transitions $v$=2--1 O(2) and $v$=3--2 O(2) are exceptionally bright relative to the ortho-lines. Extending the extraction into the G395H grating also reveals broad 3.3 and 3.4 $\mu$m PAH emission features, typically indicative of UV irradiation (Mentzer et al., in prep).


Table \ref{tab:h2_line_fluxes} summarizes the detected H$_2$ ro-vibrational transitions shown in Figure \ref{fig:fig2}, including their atomic parameters, continuum-subtracted aperture-averaged surface brightnesses, and ratios relative to the $v{=}1$--0 O(2) transition. Surface brightnesses are computed by frequency-integrating the continuum-subtracted spectra in each pixel within the extraction aperture and then averaging over all pixels. Uncertainties are obtained by propagating the local RMS noise through the spectral integration, with an additional 7\% absolute flux calibration uncertainty added in quadrature.

\begin{table*}[t]
\centering
\begin{tabular}{l c c c c c c c c c c}
\hline
Line & $\lambda$ ($\mu$m) & $(v_u,J_u)$ & $(v_l,J_l)$ & $E_u/k$ (K) & $E_l/k$ (K) & $g_u$ & $g_l$ & $A_{ul}$ (s$^{-1}$) & Brightness (erg s$^{-1}$ cm$^{-2}$ sr$^{-1}$) & Line Ratio \\
\hline
$v$=1-0 O(2) & 2.627 & (1,0) & (0,2) & 5987 & 509 & 1 & 5 & 8.53$\times10^{-7}$ & $(5.10\pm0.36)\times10^{-5}$ & $1.000$ \\
$v$=2-1 O(2) & 2.786 & (2,0) & (1,2) & 11635 & 6471 & 1 & 5 & 1.28$\times10^{-7}$ & $(1.31\pm0.10)\times10^{-5}$ & $0.252\pm0.026$ \\
$v$=3-2 O(2)  & 2.962 & (3,0) & (2,2) & 16952 & 12094 & 1 & 5 & 1.41$\times10^{-7}$ & $(5.58\pm0.50)\times10^{-6}$ & $0.106\pm0.012$ \\
$v$=1-0 S(0) & 2.223 & (1,2) & (0,0) & 6471 & 0 & 5 & 1 & 2.53$\times10^{-7}$ & $(3.07\pm0.44)\times10^{-6}$ & $0.062\pm0.010$ \\
$v$=1-0 O(3) & 2.803 & (1,1) & (0,3) & 6149 & 1015 & 9 & 21 & 4.22$\times10^{-7}$ & $(2.83\pm0.46)\times10^{-6}$ & $0.056\pm0.010$ \\
$v$=1-0 O(5) & 3.235 & (1,3) & (0,5) & 6951 & 2503 & 21 & 33 & 2.08$\times10^{-7}$ & $(2.67\pm1.48)\times10^{-6}$ & $0.051\pm0.029$ \\
$v$=1-0 S(1) & 2.122 & (1,3) & (0,1) & 6951 & 170 & 21 & 9 & 3.47$\times10^{-7}$ & $(2.09\pm0.39)\times10^{-6}$ & $0.041\pm0.008$ \\
$v$=1-0 O(4)  & 3.004 & (1,2) & (0,4) & 6471 & 1681 & 5 & 9 & 2.90$\times10^{-7}$ & $(1.26\pm0.31)\times10^{-6}$ & $0.025\pm0.006$ \\
\hline
\end{tabular}
\caption{
Observed H$_2$ ro-vibrational line surface brightnesses toward the eastern outer disk region of Oph~163131 extracted within the $0.5''$ diameter aperture shown in Figure \ref{fig:fig2}. Atomic parameters for each line, including the upper and lower ro-vibrational states ($v_u$, $J_u)$, ($v_l$, $J_l)$, temperature-equivalent energies in Kelvin ($E_u/k$) where $k$ is Boltzmann's constant, degeneracies ($g_u$, $g_l$), and Einstein-A coefficient  ($A_{ul}$)  are adopted from \citet{Roueff_H2_2019}. In the last column, line intensities are normalized to the $v$=1--0 O(2) transition.  
}
\label{tab:h2_line_fluxes}
\end{table*}

A key challenge in interpreting the excitation source of H$_2$ in disk regions is that the emergent ro-vibrational spectrum depends not only on the underlying excitation mechanism, but also on subsequent processing of the level populations. In cold, dense environments, collisions can rapidly redistribute the population among rotational levels within a given vibrational manifold before radiative decay, preferentially suppressing high-$J$ emission \citep[e.g.,][]{Black_vanDishoeck_1987,Sternberg_Dalgarno_1989}. This effect should be especially important in outer-disk gas (on the hundred au scale), where temperatures are on the order of tens of Kelvin and gas densities can reach up to $\gtrsim10^{9}  \mathrm{cm^{-3}}$ \citep[e.g.,][]{Ewine_chemistry_review_PNAS_06}. For the remainder of this paper, we refer to this excitation regime as ``vibrationally hot, rotationally cold" (or simply ``v-hot, J-cold" for short) H$_2$ emission.

The combination of (i) a very strong 1--0 O(2) line, (ii) weak but detectable 1--0 O(3) emission, (iii) extremely faint transitions originating from $J{=}2$, and (iv) the presence of bright 2--1 and 3--2 O(2) lines constitutes the central observational signature of the Oph 163131 H$_2$ spectrum, characteristic of the ``v-hot, J-cold" excitation regime. While this paper focuses on the bright outer-disk extraction shown in Figure \ref{fig:fig2}, preliminary spatially resolved analysis suggests that the low O(3)/O(2) behavior remains broadly consistent across much of the extended disk, with stronger spatial variations appearing only very close to the center. A more complete analysis of these spatial variations, together with comparisons across other edge-on sources in the JEDIce program, is analyzed in detail in follow-up work.

\section{Non-thermal excitation of ro-vibrational H$_2$ in a cold, dense gas}\label{sec:3_rovibexcitation}
In classical UV-excitation, rotational populations within a given $v$ manifold are typically distributed across multiple $J$ levels, and transitions such as $v{=}$1--0 S(0) and 1--0 O(4) are commonly detected with comparable strengths to the 1--0 O(2) line. In the CR-isolated B68 spectrum presented by \citet{Bialy2025_B68_CRH2} and listed in \citet{Neufeld2025_CRinB68_JWST} (their table 2), the measured 1--0 O(3)/O(2) and 1--0 O(4)/O(2) ratios are $\sim0.16$ and $\sim0.31$, respectively, whereas in Oph 163131 we measure even smaller values of $\sim0.06$ and $\sim0.03$ (Table 1). The similarly strong suppression of odd-J and higher-J transitions relative to O(2) therefore already suggests that standard UV excitation alone may not fully explain the observed spectrum and indicates that additional processing (e.g., efficient collisional de-excitation) occurs after excitation and before ro-vibrational radiative decay.

At the same time, the detection of bright 1--0, 2--1, and 3--2 O(2) emission demonstrates that the excitation mechanism is capable of populating $v\ge1$ vibrational levels. This requirement is impossible to satisfy through thermal excitation alone (i.e. collisional excitation by molecules at the kinetic temperature of the gas) at the low temperatures inferred for the outer disk ($\sim$30–40 K at radii beyond 200 au; \citealt{Flores_unusualanatomy_oph163131_2021,Villenave_highly_setled_Oph163131_2022}). The observed emission therefore points to a non-thermal origin for the initial excitation, followed by suppression of higher-$J$ level populations through collisions in cold, dense gas.

\subsection{Illustrative UV-only models}\label{sec:results:meudon_uv}

\begin{figure*}
\centering
\includegraphics[width=0.47\linewidth]{ 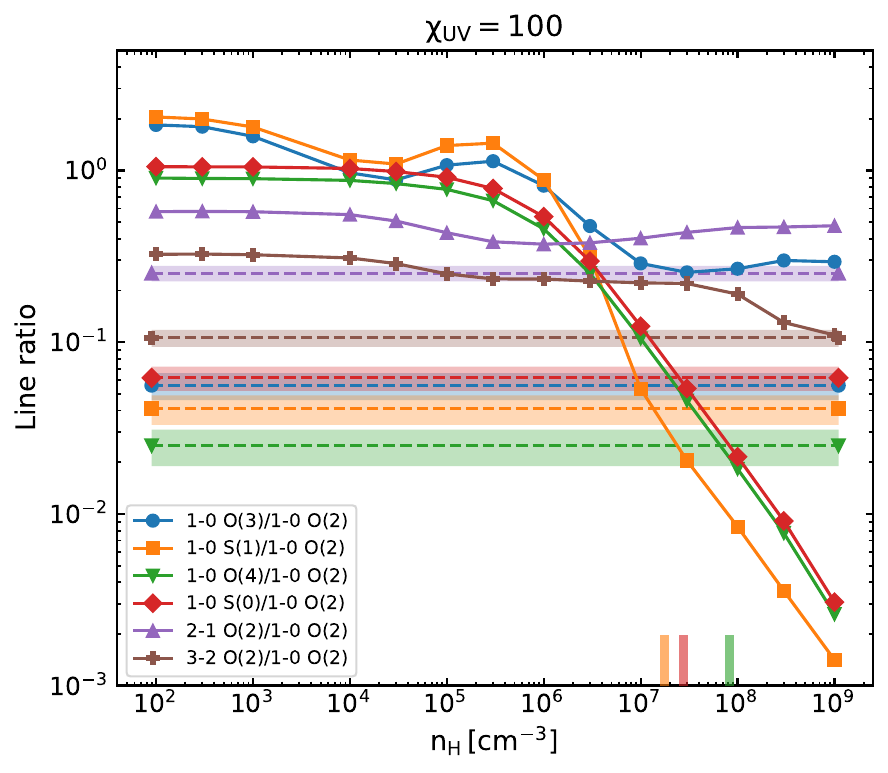}\hfill
\includegraphics[width=0.47\linewidth]{ 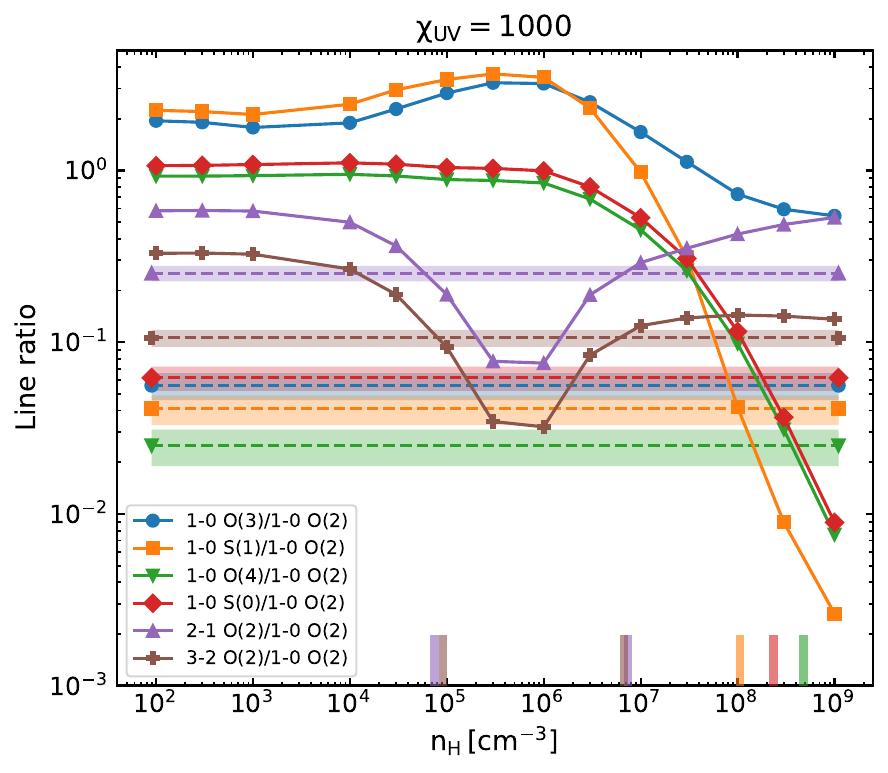}

\includegraphics[width=0.47\linewidth]{ 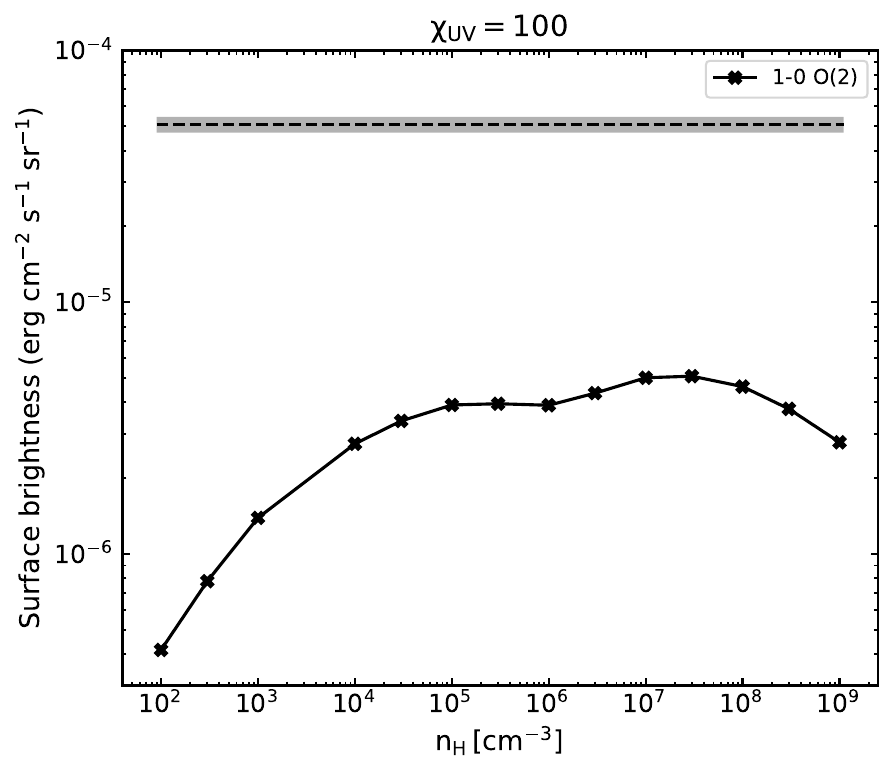}\hfill
\includegraphics[width=0.47\linewidth]{ 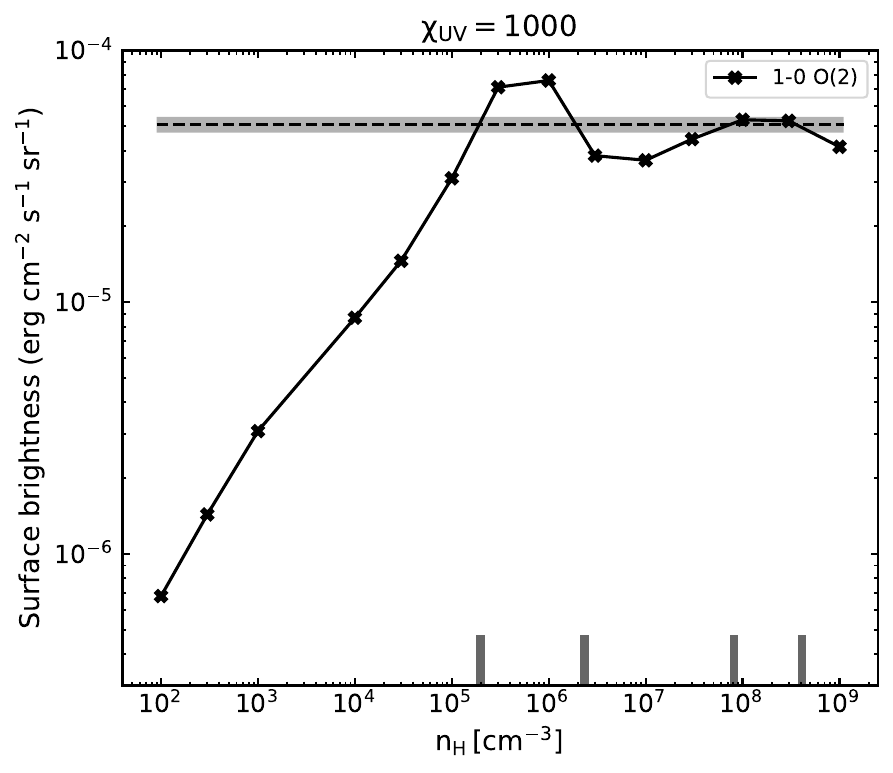}

\vspace{0.35em}

\caption{
Comparison of Meudon PDR model predictions with observed H$_2$ constraints for $\chi_{\rm UV}=100$ (left) and $1000$ (right) UV radiation fields. Top row: line ratios relative to 1--0 O(2). Horizontal dashed lines indicate observed values, shaded bands show observational uncertainties, and vertical ticks denote gas densities where the model predictions intersect the observed constraints. Bottom row: absolute surface brightness as a function of gas density for the 1-0 O(2) line.
}
\label{fig:meudon_chi100_1000}
\end{figure*}

We first explore whether ultraviolet excitation alone, operating at high enough gas densities where collisional de-excitation is efficient, can broadly reproduce the features of the observed ro-vibrational H$_2$ spectrum. To do this, we ran isochoric (constant-density) models using the Meudon PDR code, where the gas is treated as a 1D plane-parallel slab with isotropic external UV irradiation incident on the slab surfaces \citep[see Fig.~1]{LePetit_MeudonPDR_2006}.
The temperature is computed self-consistently through thermal balance and varies strongly with depth into the slab as the UV field becomes attenuated by dust and H$_2$ self-shielding, producing a thin warm surface layer (typically a few hundred to a few thousand K) above a substantially cooler interior with temperatures of tens of K.
We extend beyond the diffuse PDR regime to densities characteristic of outer-disk gas, spanning hydrogen densities $n_{\rm H}=10^{2}$--$10^{9}$ cm$^{-3}$ and normalized UV field strengths $\chi_{\rm UV}=100$ and $1000$, corresponding to 100 and 1000 times the local interstellar radiation field in the solar neighborhood as defined by Draine \citep[][]{Draine_photoelectric_heating_interstellar_gas_78, Bialy_cold_clouds_CR_detecttors_2020}.


Figure \ref{fig:meudon_chi100_1000} illustrates how increasing gas density in both UV models modifies the emergent H$_2$ line ratios relative to the 1--0 O(2) line (top panels), as well as the surface brightness of this line (bottom panels). Superposed on the models are the observed vibrational O(2) line brightnesses and ratios from the spectrum in Figure \ref{fig:fig2}.

The models broadly reproduce several key aspects of the observed spectrum, including the qualitative ordering of the O(2) lines, with the $v{=}1$--0 transition brighter than $v{=}2$--1, which in turn is brighter than $v{=}3$--2 (purple and brown curves in the upper panels of Figure \ref{fig:meudon_chi100_1000}). Increasing the gas density progressively suppresses high-$J$ emission within the $v{=}1$ manifold, reducing the strength of transitions originating from $J{=}2$ (1--0 S(0) and O(4); red and green curves) and $J{=}3$ (1--0 S(1); orange curve). These lines fall to or below the observed ratio levels at densities of $\sim10^{7}$–$10^{8}\mathrm{cm^{-3}}$ for the $\chi_{\rm UV}=100$ model and $\sim10^{8}$–$10^{9}\mathrm{cm^{-3}}$ for $\chi_{\rm UV}=1000$, as indicated by the vertical tick at the bottom of the upper panels. Together, these models demonstrate how UV excitation to $v\ge2$ combined with collisional de-excitation of higher-$J$ levels within the $v{=}1$ manifold naturally produces vibrationally hot, but rotationally cold H$_2$ emission.

Quantitatively, however, there are notable discrepancies between the model predictions and the observations. In particular, the $\chi_{\rm UV}{=}100$ model overpredicts the intensities of, both, the $v{=}3$–2 and $v{=}2$–1 O(2) lines relative to the 1–0 O(2) line by approximately a factor of two, especially in the $n_{\rm H}\sim10^{7}$–$10^{8}\mathrm{cm^{-3}}$ regime where collisions suppress the 1–0 S(0), O(4), and S(1) lines to the observed levels relative to 1–0 O(2). A further discrepancy is that the model overpredicts the intensity of the 1–0 O(3) line relative to 1–0 O(2) by approximately a factor of $\sim$4.5 at these densities (blue lines in the upper-left panel. The 1–0 O(3) transition arises from the $v{=}1$, $J{=}1$ level, which is the lowest-energy odd-$J$ state in the manifold and is therefore not strongly affected by collisional de-excitation at high densities. As a result, the excess O(3) emission cannot be mitigated by collisional depopulation alone and instead points to additional effects influencing the excitation conditions (see further discussions below). Finally, the model underpredicts the absolute surface brightness of the 1--0 O(2) line by roughly an order of magnitude (bottom-left panel). This discrepancy may plausibly be alleviated by a combination of geometric and dust-evolution effects. In particular, the nearly edge-on viewing geometry may increase the effective path length through the shallow UV-pumped outer disk layers relative to the perpendicular sightline assumed in the plane-parallel slab models, while grain growth and strong dust settling in the Oph 163131 disk may further reduce the abundance of relatively small grains responsible for near-IR extinction, allowing emission from a larger effective H$_2$ column to escape.

The $\chi_{UV}$=1000 model matches the observed spectra better than the $\chi_{UV}$=100 model in some aspects, but worse in others. Specifically,  it reproduces the observed $v$=2-1/1-0 and 3-2/1-0 O(2) line ratios at densities around  n$_H\sim 10^5$ and 10$^7$ cm$^{-3}$ (see the purple and brown curves in the top-right panel).  It also roughly matches the observed 1-0 O(2) line intensity at the high end of the explored density range (see the bottom-right panel). However, it exacerbates the discrepancy in the 1-0 O(3) line intensity, over-predicting it by an order of magnitude (rather than a factor of $\sim$4.5 in the $\chi_{UV}=$100 case, blue curves in top panels). Because this ratio probes the relative population of $J=1$ and $J=0$ levels within the $v=1$ manifold, it provides a particularly strong constraint on the ortho-to-para ratio and hence the excitation mechanism. In other words, the gas within the line of sight emitting the observed H$_2$ spectrum should be relatively cold and the $v=1$, $J=1$ population is initially low compared to the $v=1$, $J=0$ ground state, which is inconsistent with the heating that is expected from these UV models, especially the $\chi_{UV}$=1000 model.

As previously mentioned, the CO($J{=}2$–1) emission observed with ALMA indicates characteristic gas temperatures of $T\sim30$–40 K \citep{Flores_unusualanatomy_oph163131_2021, Villenave_highly_setled_Oph163131_2022} along the same lines of sight, reinforcing the interpretation that at least part of the molecular gas associated with the line-of-sight H$_2$ emission is cold. A stratified geometry is then plausible, in which higher-$J$ and ortho lines (e.g., 1–0 O(3)) arise primarily from a warmer UV-irradiated component in the outer disk along the line of sight through the midplane, while the dominant 1–0 O(2) emission traces both this component and colder, denser CO-emitting gas deeper in the disk (see the cartoon and associated discussion in Section \ref{sec:discusccion:volume_surface_excitation} below).

\subsection{The case for a cosmic-ray contribution}\label{sec:results:cr_modeling}

A key characteristic of the outer-disk H$_2$ spectrum in Oph 163131 is the dominance of the 1–0 O(2) line relative to higher-$J$ transitions (Figure \ref{fig:fig2}), a feature that closely resembles the CR–excited H$_2$ spectrum observed toward the starless core B68 \citep[][see their Figure 2]{Bialy2025_B68_CRH2}. This similarity makes CR excitation a plausible contributor for the observed emission. It also alleviates several qualitative and quantitative discrepancies encountered in the UV-only models discussed in Section \ref{sec:results:meudon_uv}. In particular, CRs are expected to penetrate deeper into the disk than UV photons and excite H$_2$ over a larger volume that is more consistent with the observed morphology of the O(2) emission (see the first panel of Figure \ref{fig:fig1}). Moreover, CR excitation does not require substantial gas heating, yielding H$_2$ emission from gas that remains relatively cool, as in the B68 case, and more consistent with the CO-inferred temperatures of $\sim$30–40 K in Oph 163131.

As emphasized by \citet{Bialy2025_B68_CRH2}, such low gas temperatures naturally suppress the population of the $v{=}0$, $J{=}1$ level (with an upper energy of $\sim$171 K) relative to the ground state $v{=}0$, $J{=}0$. This leads to a reduced ortho-H$_2$ emission relative to para-H$_2$, thus alleviating one of the major quantitative discrepancies of the UV-only models, namely the overprediction of the ortho 1–0 O(3) line strength relative to the para 1–0 O(2) line (see the blue curves in the upper panels of Figure \ref{fig:meudon_chi100_1000}). Indeed, in the limit of very cold gas, CR-only models tend to underpredict the 1–0 O(3) emission altogether, indicating that at least a modest contribution from warmer gas is required. This is most plausibly provided by UV-heated gas in the outer disk, as similarly inferred for the starless core B68, where the more externally UV-irradiated gas dominates the 1–0 O(3) emission relative to the deeper, more shielded gas (see the schematic and discussion in Section \ref{sec:discussion}).

Furthermore, a CR contribution provides a natural way to reconcile another key discrepancy in the $\chi_{\rm UV}=100$ model, namely the factor-of-two overprediction of the $v{=}2$--1 and $v{=}3$--2 O(2) lines relative to the $v{=}1$--0 transition (see Section~\ref{sec:results:meudon_uv}). Secondary electrons produced by CR ionization excite H$_2$ efficiently into the $v{=}1$ manifold, while contributing only weakly to $v\ge2$. For example, \citet{Gredel_infrared_H2_Xrays_1995} show that for $\sim$20~eV electrons incident on H$_2(J{=}0)$, the relative excitation rates into $v{=}1$, $v{=}2$, and $v{=}3$ are $1.000{:}0.083{:}0.001$. This is in broad agreement with the relative excitation to $v=1$, $v=2$, and $v=3$ computed by integrating the secondary electron spectra from \citet{Padovani2022_CR_H2} over the recently revised excitation cross sections by electron impact \citep[][Padovani et al. in prep.]{Scarlett_H2collisionsI_2021,Scarlett_H2_elastic_scattering_rot_exc_2023}. As a result, adding a CR-driven component preferentially enhances the 1--0 O(2) intensity, while leaving the intensity of higher-$v$ O(2) lines largely unchanged.

In this sense, CR excitation acts to dilute the UV-pumped contribution to the $v{=}2$--1 and $v{=}3$--2 O(2) ratios: if roughly half of the observed 1--0 O(2) emission arises from CR excitation, the predicted $v{=}2$--1/1--0 and $v{=}3$--2/1--0 ratios are reduced by approximately the factor of two required to match the observations. This motivates adopting a mixed excitation scenario, in which UV photons dominate the production of $v\ge2$ emission, while CRs contribute to the $v{=}1$ population at a level comparable to UV photons. In this case, the CR excitation contributes approximately half of the observed 1-0 O(2) surface brightness, $I_{\rm CR} \approx I_{\rm obs}/2$. We note, however, that while the specific example of CR contributing to half of the observed 1-0 O(2) intensity can remove the factor-of-two overprediction of the higher $v$ 3-2 and 2-1 O(2) lines relative to the 1-0 O(2) line in the $\chi=100$ model, the overprediction factors (and thus the CR contributions needed to correct them) may be model-dependent. The corresponding CR ionization rate inferred here should be treated as an order-of-magnitude illustration rather than a precise inference. Nevertheless, using the formalism of \cite{Bialy2025_B68_CRH2}, which applies in the purely radiative limit (i.e., without collisional redistribution), the corresponding order-of-magnitude CR ionization rate estimate is then

\begin{equation}\label{eq:eq1}
    \zeta_{\rm CR} =
\frac{I_{\rm CR}}
{(1/4\pi) b_u \alpha_{ul} E_{ul} g N(\rm H_2)}
\approx
4\times10^{-14}\ \rm s^{-1},
\end{equation}
where we adopt $I_{\rm CR}\approx2.5\times10^{-5}$ erg s$^{-1}$ cm$^{-2}$ sr$^{-1}$, $\lambda = 2.627 \mu$m for the O(2) transition ($E_{ul}=7.6\times10^{-13}$ erg), a branching ratio $\alpha_{ul}=A_{ul}/A_{\rm tot}\simeq1.0$, and a CR excitation probability $b_{(v=1,J=0)}=1.57$ from Table~1 of \citet{Bialy2025_B68_CRH2}, and the same effective emitting column, $gN(\mathrm{H}_2)\approx6.5\times10^{21}$ cm$^{-2}$, as adopted by \citet{Bialy2025_B68_CRH2} for the starless core B68. However, the effective emitting column of the H$_2$ gas in the edge-on disk of Oph 163131 may differ from that of B68, owing to differences in dust properties. For example, extreme dust settling can greatly deplete the dust-to-gas mass ratio in the H$_2$-emitting regions above and below the very thin dust layer probed by ALMA continuum observations. The resulting reduction in dust opacity (cross-section per gram of gas) allows H$_2$ emission from a much larger effective column, $gN(H_2)$, to reach the observer. In such cases, the effective emitting column may exceed the fiducial value adopted here, perhaps by as much as an order of magnitude or even more (the exact factor is highly uncertain), thereby reducing the cosmic-ray ionization rate required to reproduce the observed 1-0 O(2) brightness by a comparable factor.

Additional reductions are possible when we consider collisional de-excitation, as we did for UV (Section \ref{sec:results:meudon_uv}). At the relatively high gas densities expected in the outer disk (compared to the starless core, B68), collisional de-excitation will efficiently redistribute population toward the $J=0$ level, enhancing the O(2) line intensity for a given CR ionization rate. As a result, fewer CR ionizations are required to reproduce a given O(2) surface brightness than in the low-density, purely radiative limit assumed in Equation \ref{eq:eq1}, reducing the inferred ionization rate by a factor of roughly two (see Appendix \ref{app:cr_collisional_processing}). Both the illustrative CR redistribution calculation in Appendix \ref{app:cr_collisional_processing} and the UV PDR models in Section \ref{sec:results:meudon_uv} suggest that efficient collisional redistribution within the $v=1$ manifold becomes important at densities broadly of order $n_{\rm H}\sim10^{7}$--$10^{8},\mathrm{cm^{-3}}$ in Oph 163131, although the precise threshold depends sensitively on temperature and collider rates. Corrections for \textit{foreground} extinction could modestly increase this estimate, 
though the magnitude of this effect remains uncertain. Even without explicitly accounting for extinction, and allowing for the reduction effects discussed above, the inferred ionization rate remains elevated ($\zeta_{\rm CR}\gtrsim10^{-15}$ s$^{-1}$) compared to typical values inferred for protoplanetary disks, which range from $\lesssim10^{-19}$ s$^{-1}$ for the TW Hydra disk \cite{cleeves_xray_v_CR_rates_TW_Hya_2015} to a few times  $10^{-17}$ s$^{-1}$ for the outer disk of IM Lup \citep{Seifert_CR_gradient_IM_lup_disk_2021}, based on modeling of molecular ions such as HCO$^+$.

\section{Discussion}\label{sec:discussion}

The analysis above shows that, both, UV photons and CRs are required to explain the excitation of ro-vibrational H$_2$ in our outer disk spectrum. We also show that collisions are needed to explain the suppression of higher-J lines, which together with these non-thermal excitation conditions, broadly explain the vibrationally hot, rotationally cold H$_2$ pattern. Here, we summarize our interpretation of the excitation conditions and then briefly discuss broader implications in regard to disk chemistry and evolution. 

\subsection{Stratified excitation of ro-vibrational H$_2$ in disks}
\label{sec:discusccion:volume_surface_excitation}

\begin{figure*}
    \centering
    \includegraphics[width=0.99\linewidth, trim={0cm 0cm 0cm 0cm},clip]{ 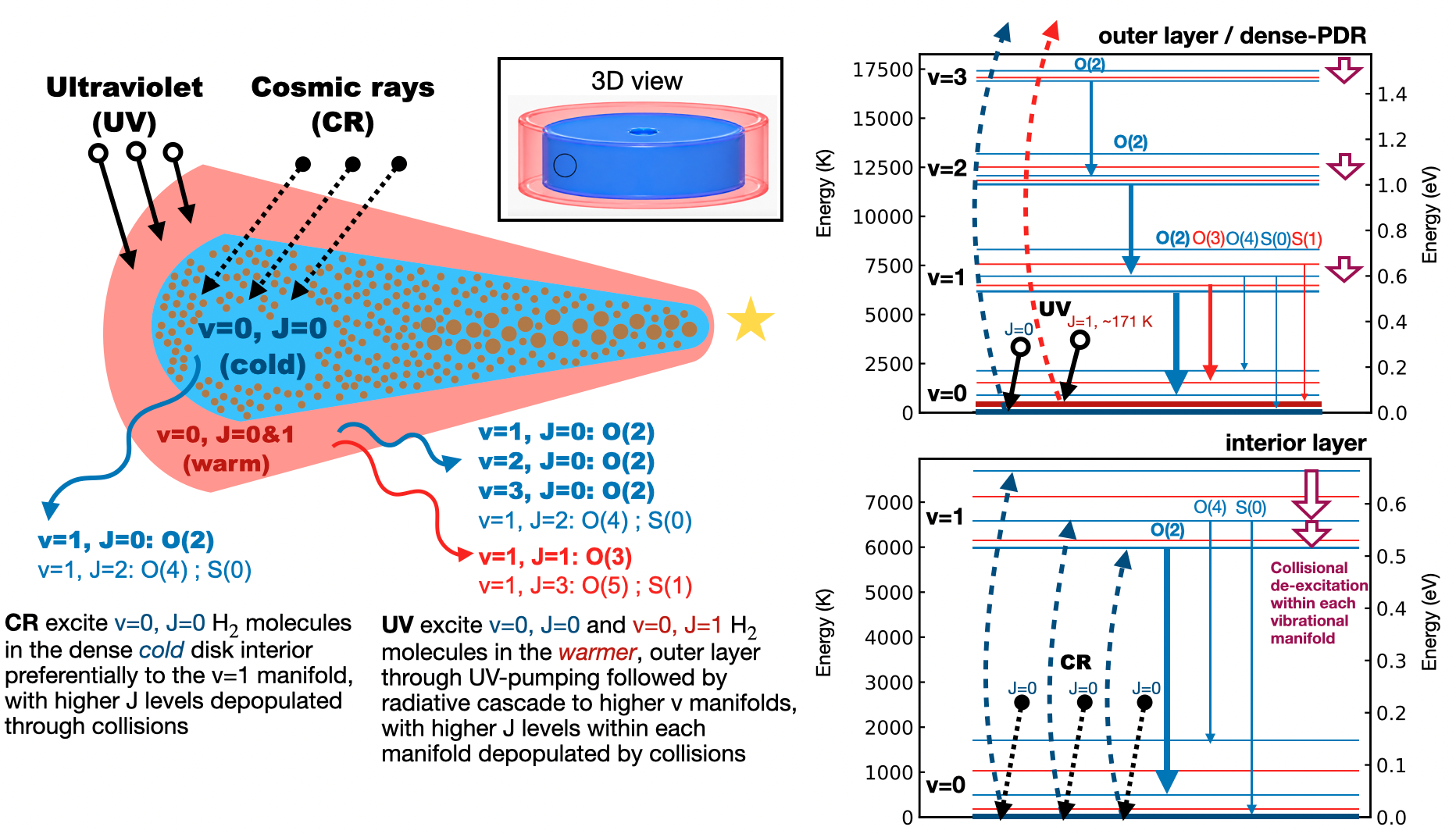}
\caption{
Schematic illustration of a warm irradiated outer component and more interior excitation of ro-vibrational H$_2$. Left: Disk meridional cross-section showing UV excitation confined to a warm irradiated layer and CR excitation penetrating more deeply into the molecular disk. The inset above the cross-section shows the corresponding simplified 3D axisymmetric geometry obtained by rotating the 2D structure about the stellar axis. Right: Corresponding vibrational–rotational energy-level diagrams. In both cases, excitation produces vibrationally hot, yet rotationally cold H$_2$ due to collisional de-excitation. The outer irradiated component (top) enables emission from higher (v=2 and 3) emission for O(2) and significant emission from ortho-lines, particularly 1-0 O(3), while the interior component (bottom) contributes mostly low-J para-lines from the v=1 manifold, particularly O(2) and to a lesser extent, O(4) and S(0). The observer is assumed to view the disk nearly edge-on, with a sightline passing through both the warmer UV-irradiated outer component and the CR-excited deeper interior.
}


\label{fig:fig_schematic}

\end{figure*}

Our interpretation of the observed ro-vibrational H$_2$ emission in Oph 163131 is summarized in Figure \ref{fig:fig_schematic}. UV photons (open circle, solid arrows) capable of H$_2$ excitation penetrate (and warm up) only a relatively low column density region of the outer disk regions due to attenuation by dust or H$_2$ self-shielding. In this warm and dense PDR layer (red-shaded in the cartoon), the ground-state population includes both $v=0, J=0$ and $J=1$, and UV excitation populates high vibrational states from UV pumping followed by radiative cascade, allowing emission from multiple vibrational and rotational levels, including higher-$J$ para and ortho transitions (blue and red arrows, respectively). At the same time, the gas densities are sufficient for collisional redistribution within each vibrational manifold, which suppresses higher-$J$ emission and favors low-$J$ lines, leaving primarily O(2) ($J=0$) and O(3) ($J=1$) emission from the outer disk irradiated component (indicated by thicker arrows and more prominent labels throughout Figure~\ref{fig:fig_schematic}).

In contrast, CRs (solid dotted arrows in Figure \ref{fig:fig_schematic}) penetrate deeper into the disk and excite H$_2$ within cold, dense molecular gas that samples a more interior layer. In this regime, most of the gas resides in $v=0$, $J=0$ and is primarily excited to even para levels within the $v=1$ manifold (see Section \ref{sec:results:cr_modeling}). At these higher densities, collisional redistribution further suppresses higher-$J$ emission, leading to strong low-$J$ para emission, particularly the $v$=1-0 O(2) line. The other para levels (e.g., $v=1$, $J=2$) within the $v=1$ manifold are still populated by CR excitation, but collisional depopulation leads to weaker (but still detectable) 1-0 S(0) and O(4) lines relative to 1-0 O(2) compared to the lower-density starless core B68 \citep[][see Appendix \ref{app:cr_collisional_processing} for more details]{Bialy2025_B68_CRH2} 

The energy-level diagrams on the right side of Figure \ref{fig:fig_schematic} summarize these excitation regimes in terms of their vibrational and rotational populations. The upper diagram represents the warm irradiated outer component (dense-PDR) case, where UV excitation populates high vibrational states from both $J=0$ and $J=1$, while collisional redistribution favors emission from the lowest rotational levels. The lower diagram illustrates the more interior, penetrating case, in which CRs excite predominantly cold $v=0$, $J=0$ gas preferentially to $v=1$ (rather than higher $v$ manifolds) and collisions similarly drive the emission toward low-$J$ para transitions. In both regimes, the result is vibrationally excited, yet rotationally cold H$_2$.

\vspace{-0.5em} 
\subsection{Protostellar and external sources of excitation}
\label{sec:discusccion:internal_v_external_excitation}

The presence of far-UV irradiation in the environment of Oph 163131 is supported by multiple independent observational tracers. The detection of strong 3.3 and 3.4 $\mu$m PAH features, together with vibrationally excited H$_2$ emission ($v>1$), demonstrates that far-UV photons reach the disk surface (Mentzer et al., in prep.). In addition, ALMA observations show that the gas temperature inferred from CO increases with radius along the midplane, reaching $\sim$30–40 K in the outer disk \citep{Flores_unusualanatomy_oph163131_2021}. In a highly inclined, dust-settled disk, such temperatures are difficult to reconcile with stellar irradiation alone and instead point to heating by an external far-UV radiation field with intensity on the order of 100 times solar ($\chi_{UV}\approx100$). Consistent with this picture, JWST/MIRI imaging reveals large-scale PAH and H$_2$ emission extending beyond the disk \citep{Villenave_JWST_mid_IR_oph163131_2024}, inferred to originate from an externally influenced UV environment.

Similar UV field strengths have been inferred elsewhere in the $\rho$ Ophiuchi cloud, where localized external radiation fields spanning $\chi_{UV} \sim 10^2$–$10^3$ have been measured from JWST observations of irradiated outflows and surrounding material \citep{Skretas_uv_irrad_outflows_oph_2025}. Oph 163131 lies eastward of and offset from the main $\rho$ Ophiuchi cloud star-forming core and to the north of the Upper Scorpius association. Within this broader context, it is plausible that Oph 163131 experiences an elevated UV flux, under the influence of nearby shocks or from more distant luminous stars whose radiation permeates the cloud. However, no nearby O- or B-type stars are directly associated with Oph 163131, and the nearest bright Upper Sco source (i Sco; B2.5) lies $\sim$37$'$ away, limiting its likely influence on the disk. Low-resolution H$\alpha$ maps from the Southern H-Alpha Sky Survey Atlas  \citep[SHASSA;][]{Gaustad_SHASSA_2001} show a low level of diffuse emission to the south and southeast of Oph 163131, spatially coincident with Haro 1-16 at a projected separation of $\sim$1$'$, although its impact on the disk excitation remains uncertain.

The elevated CR ionization rates inferred here, $\zeta \sim 10^{-15}$–$10^{-14}$ s$^{-1}$, exceed typical values measured in the Galactic ISM and starless cores such as B68 ($\sim10^{-16}$ s$^{-1}$; \citealt{Bialy2025_B68_CRH2}), indicating that the typical diffuse Galactic CR background is unlikely to be sufficient. Instead, these values suggest either a local enhancement within the surrounding cloud or an additional contribution associated with the protostellar environment, potentially including shocks or energetic particles generated during accretion and outflow activity \citep[e.g.,][]{Padovani_CR_accel_protostars_2015, Padovani_forges_of_CR_protostars_2016, Brunn_ionization_t_tauri_in_situ_particles_mag_reconnection_2023}. 

\subsection{High cosmic-ray rates in disks}

High CR ionization rates comparable to those explored here have been reported in a small number of dense star-forming environments. For example, \citet{Cabedo2023} inferred $\zeta\sim10^{-15}$–$10^{-14}$ s$^{-1}$ in the inner envelope of the deeply embedded protostar B335, based on enhanced emission from molecular ions such as HCO$^+$, DCO$^+$, and N$_2$H$^+$. Similarly elevated values have been inferred in OMC2 FIR4 \citep{Ceccarelli2014, Fontani_OMC2_FIR4_2017}. In these cases, the high ionization rates were interpreted as evidence for locally accelerated energetic particles associated with protostellar activity.

The ionization rates explored for Oph  163131, therefore, lie in a regime that is unusual, but not without precedent. A key distinction is that, unlike B335, where elevated $\zeta$ characterizes the inner infalling envelope, the H$_2$ emission in Oph  163131 arises in the extended outer disk. Whether energetic particles in this system are primarily external in origin, locally accelerated near the protostar, or guided into the disk by magnetic field structures remains uncertain. Magnetic field strengths and geometries play a critical role in regulating particle propagation, but are poorly constrained observationally.

If the elevated ionization rates explored here are confirmed, they would have important consequences for the chemistry and dynamics of the outer disk. For example, enhanced ionization increases the abundance of molecular ions and of He$^{+}$, which can efficiently destroy CO through gas-phase reactions and redistribute carbon into CO$_2$, hydrocarbons, and more complex organics via subsequent grain-surface chemistry \citep[e.g.,][]{Schwarz2018}. Such processes have been invoked to explain CO depletion in disks, although the extent to which they operate in Oph 163131 remains uncertain.

At the same time, increased ionization enhances the coupling between the gas and magnetic fields by increasing the abundance of charged species, including molecular ions, PAH ions, and charged small grains. This can modify the strength of the coupling between magnetic fields and the bulk neutral gas, with potential consequences for angular momentum transport and the launching of magnetized disk winds. 

\section{Conclusion}
\label{sec:conclusion}

JWST/NIRSpec IFU observations of the edge-on disk Oph 163131 reveal an unusual H$_2$ ro-vibrational spectrum, in which emission in the outer disk regions is systematically dominated by the O(2) transitions across multiple vibrational manifolds, including 1–0, 2–1, and 3–2. In each case, the lowest-$J$ para line is significantly brighter than adjacent ortho and higher-$J$ transitions, while higher-$J$ emission within the $v=1$ manifold is strongly suppressed. This repeated hierarchy is highly atypical of standard UV or shock-excited spectra and indicates a vibrationally excited, yet rotationally cold H$_2$ population.

We show that this hierarchy can broadly be explained when non-thermal excitation operates in dense outer-disk gas, where collisions efficiently redistribute level populations within a vibrational manifold prior to radiative decay. Once this regime is reached, the emergent spectrum becomes vibrationally hot, but rotationally cold, and line emission from the same ``$v$", but different ``$J$" levels is only weakly sensitive to the identity of the initial excitation mechanism. UV-only, dense PDR models can reproduce some aspects of the observed spectrum, particularly the domination of the O(2) lines and the ordering of these lines in brightness, but do not reproduce the line ratios quantitatively, particularly when the observed lines from the $v=1$ level (such as O(3)) are taken into account.

Compared to UV, the CRs have the advantage of readily populating the $v$=1 levels without overheating the gas and thus alleviating the problem of overpredicting the para-to-ortho line ratios, and penetrating deeper into the disk, as required by the O(2) emission morphology. However, they underpredict the 3-2 and 2-1 O(2) line intensity relative to  1-0 O(2) line (opposite to the UV case), and thus are unlikely to provide an explanation of the observed spectra \textit{alone}. We conclude that the most plausible scenario is that both contribute, as in the case of the starless core B68 and illustrated schematically in Figure \ref{fig:fig_schematic}, although details remain to be fully quantified in future investigations.

More generally, this study demonstrates that post-excitation collisional processing can dominate the appearance of ro-vibrational H$_2$ spectra in dense outer disks, reshaping non-thermal excitation signatures without necessarily erasing them when the local density and temperature are well constrained. Ro-vibrational H$_2$ emission can therefore serve as a probe of both irradiation and cosmic-ray ionization in disks, with important implications for disk ionization, chemistry, and magnetic coupling, provided that collisional de-excitation in the dense gas from which the emission originates is explicitly accounted for.  

\begin{acknowledgements} This work is based on observations made with the James Webb Space Telescope with data obtained from the Mikulski Archive for Space Telescopes (MAST) at the Space Telescope Science Institute under NASA contracts NAS 5-26555 and NAS 5-03127. Support for Program number 5299 was provided by NASA
through a grant from the Space Telescope Science Institute,
which is operated by the Association of Universities for
Research in Astronomy, Inc., under NASA contract NAS
5-03127. The specific observations analyzed in this work can be accessed via \dataset[doi: 10.17909/hmaf-9z97]{https://doi.org/10.17909/hmaf-9z97}. This work also made use of observations from the Atacama Large Millimeter Array (ALMA), (2016.1.00771.S, 2018.1.00958.S). ALMA is a partnership of ESO (representing its member states), NSF (USA) and NINS (Japan), together with NRC (Canada), NSTC and ASIAA (Taiwan), and KASI (Republic of Korea), in cooperation with the Republic of Chile. The Joint ALMA Observatory is operated by ESO, AUI/NRAO and NAOJ. PDR models published in this paper have been produced with the Meudon PDR code (\citet{LePetit_MeudonPDR_2006}, version PDR 7.0 released in 2024, \href{http://ism.obspm.fr}{http://ism.obspm.fr}). We thank Ted Bergin for helpful discussions on the effects of high CR ionization on CO depletion. KA and ZYL are supported in part by NASA grants (80NSSC20K0533, JWST-GO-02104, and JWST-GO-08872) and NSF AST-2308199. KA acknowledges support by an NRAO ALMA SOS award. Y.-L.Y. acknowledges support from Grant-in-Aid from the Ministry of Education, Culture, Sports, Science, and Technology of Japan (20H05844 and 25H00676). M.N.D.'s work is funded by the European Union. Views and opinions expressed are however those of the author(s) only and do not necessarily reflect those of the European Union or the European Research Council Executive Agency. Neither the European Union nor the granting authority can be held responsible for them. This work is supported by ERC grant PSII (DOI: 10.3030/101230593). M.P. acknowledges the INAF grant 2023 MERCATOR (``MultiwavelEngth signatuRes of Cosmic rAys in sTar-fOrming Regions'') and 
the INAF grant 2024 ENERGIA (``ExploriNg low-Energy cosmic Rays throuGh theoretical InvestigAtions at INAF''). D.A.N. was supported by a JWST Theory Grant, JWST-AR-06829. 
JAN and ED acknowledge support from the Thematic Action `Physique et Chimie du Milieu Interstellaire' (PCMI) of INSU Programme National `Astro', with contributions from CNRS Physique \& CNRS Chimie, CEA, and CNES. 
D.H. is supported by the Ministry of Education of Taiwan (Center for Informatics and Computation in Astronomy grant and grant number 110J0353I9) and the National Science and Technology Council, Taiwan (Grant NSTC111-2112-M-007-014-MY3, NSTC113-2639-M-A49-002-ASP, and NSTC113-2112-M-007-027). J.C.S. was supported by the Heising-Simons Foundation through a 51 Pegasi b Fellowship.
\end{acknowledgements}

\bibliography{jwst}
\bibliographystyle{aasjournalv7}

\appendix


\section{Scattered-light continuum and line-to-continuum morphology}\label{App:continuum_line_to_continuum_maps}

\begin{figure}
    \centering
    \includegraphics[width=0.99\linewidth, trim=2cm 1.5cm 2cm 2cm, clip]{ 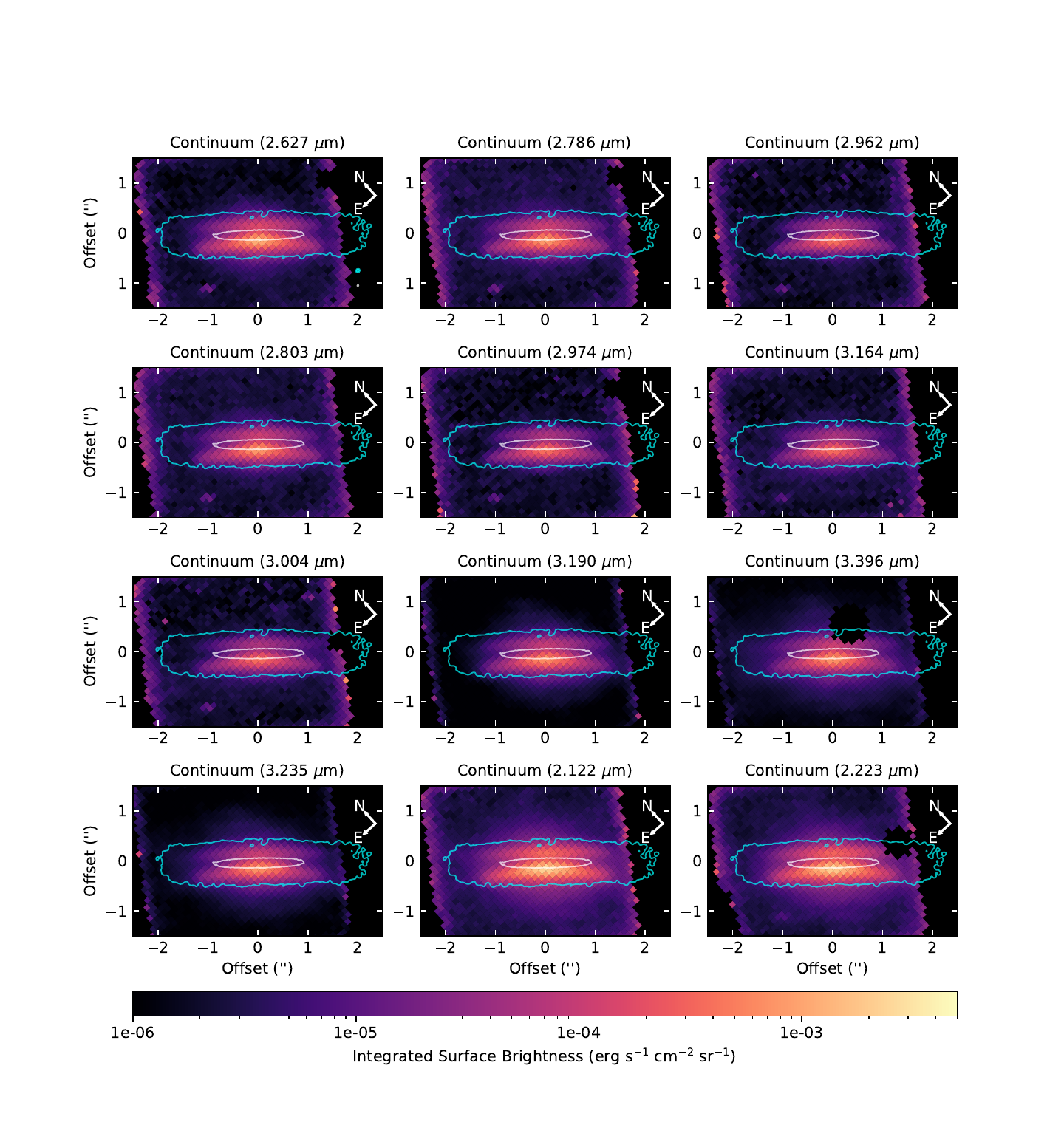}
    \vspace{-0.5cm}

    \caption{
Continuum surface-brightness maps at the wavelengths of the H$_2$ ro-vibrational transitions shown in Figure \ref{fig:fig1}, derived from the local baseline fits used in the line-flux measurements. Panels are arranged following the same row and column ordering as the line maps, with north--east arrows rotated by $40^{\circ}$ and a $250$ AU scale bar shown in each panel.
}
\label{fig:fig_app1}
\end{figure}

\begin{figure}
    \centering
    \includegraphics[width=0.99\linewidth, trim=2cm 1.5cm 2cm 2cm, clip]{ 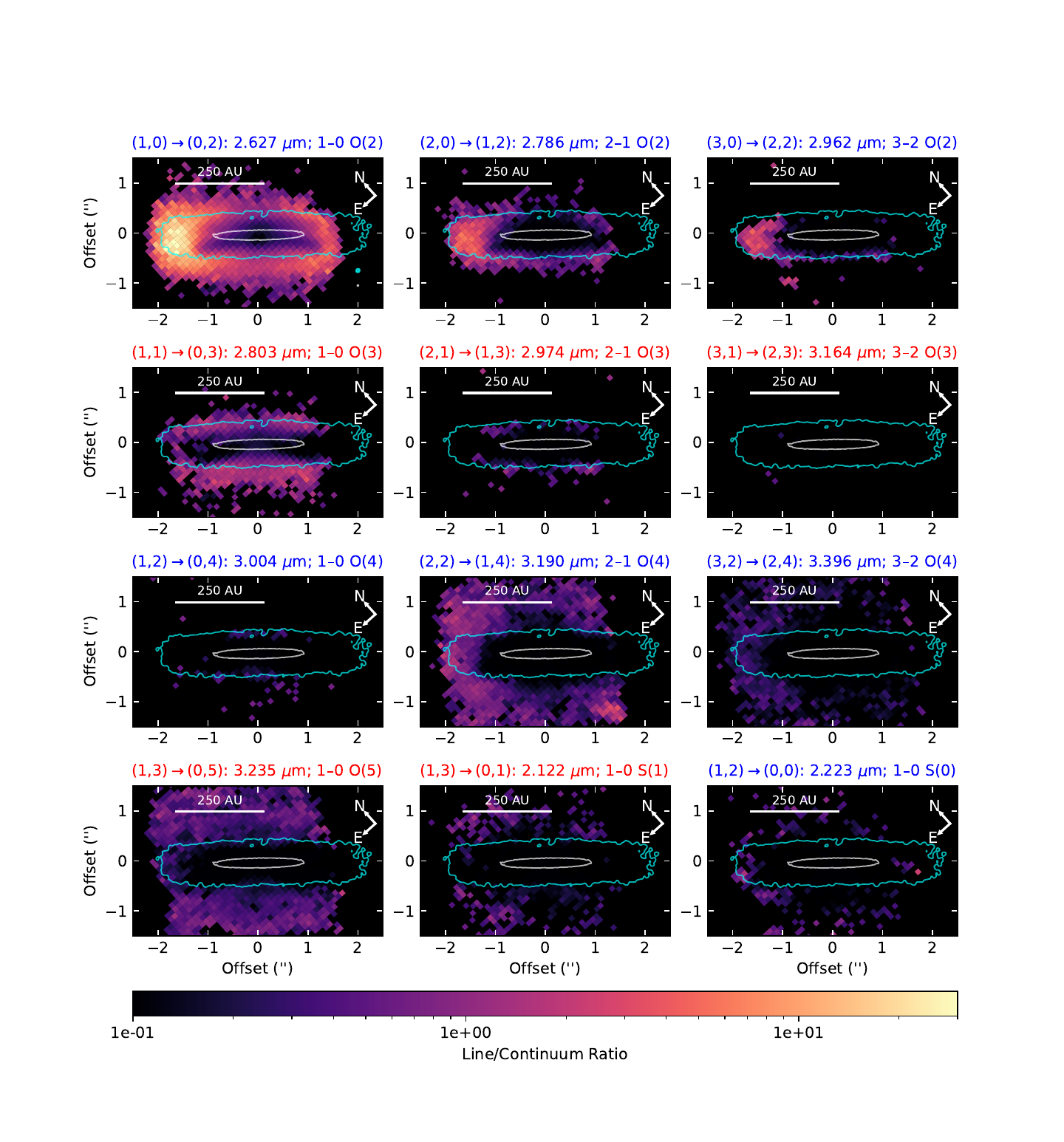}
    \vspace{-0.5cm}

    \caption{
Maps of the ratio between integrated H$_2$ line surface brightness and the corresponding integrated continuum at the same wavelength, computed over identical frequency windows. The layout and orientation match those of the line and continuum maps, emphasizing spatial variations in line excitation relative to scattered light across the disk surface.
}
\label{fig:fig_app2}

\end{figure}

To place the different H$_2$ line morphologies (Figure \ref{fig:fig1}) in context, we can also examine the underlying near-infrared continuum at the same wavelengths, which is dominated by scattered stellar light from the disk surface in this edge-on system. For each line, the local continuum is estimated via a linear fit to line-free channels and integrated over the \textit{same} frequency window (and method) used for the line measurement, yielding spatially resolved continuum surface-brightness maps (Figure \ref{fig:fig_app1}) in the same order as Figure \ref{fig:fig1}. These maps trace a bipolar reflection nebula. To isolate regions where line emission is enhanced relative to scattered light, we also show the line-to-continuum ratio maps by dividing the line surface brightness (Figure \ref{fig:fig1}) by the continuum (Figure \ref{fig:fig_app1}) at each wavelength (Figure \ref{fig:fig_app2}).

\section{Impact of Collisional De-excitation on CR-excited H$_2$}
\label{app:cr_collisional_processing}

In the main text, we argue that collisional redistribution within individual vibrational manifolds plays a central role in shaping the observed H$_2$ ro-vibrational spectrum in Oph 163131. Here, we present a simplified, worked example illustrating how this process operates in the case of CR excitation. This calculation is not intended as a complete CR excitation model, but rather as a demonstration of how collisions in cold, dense gas suppress higher-$J$ populations even in the case of CRs, helping to explain why ratios of S(0) and O(4) lines to O(2) are significantly lower than predicted in the CR model in \citet{Bialy2025_B68_CRH2}.

We consider steady-state populations in the two lowest para-H$_2$ rotational
sublevels of the first vibrationally excited state, ($v=1$, $J=0$) and ($v=1$, $J=2$). Let $n_{0,0}$ denote the population of ground-state para-H$_2$, and $n_{1,0}$ and $n_{1,2}$ the populations of ($v=1$, $J=0$) and ($v=1$, $J=2$). We define fractional populations relative to the ground state as $f_{10} \equiv n_{1,0}/n_{0,0}$ and $f_{12} \equiv n_{1,2}/n_{0,0}$. Population of the $v=1$ levels is driven by non-thermal pumping (CR secondary electrons in this particular example), parameterized by rates $P_{10}$ and $P_{12}$ into ($v=1$, $J=0$) and ($v=1$, $J=2$), respectively.

Radiative decay from both levels occurs at a rate $ A \equiv A_{1\rightarrow0} \simeq 8.5\times10^{-7}$ s$^{-1}$ (see Table \ref{tab:h2_line_fluxes}), corresponding to a radiative lifetime of $A^{-1}\simeq1.2\times10^{6}$ s. Collisional coupling within the $v=1$ manifold is described by the de-excitation rate coefficient $q_1$ for ($v=1$, $J=2$)$\rightarrow$($v=1$ ,$J=0$) and the corresponding excitation coefficient $q_2$. At $T=100$ K, we adopt $q_1 = 2.2\times10^{-13}$  cm$^{3}$ s$^{-1}$ from \citet{Flower_Roueff_H2collisions_98}, and impose detailed balance, $q_2(T) = 5  q_1  \exp(-484 \text{ K}/T)$, giving $q_2\simeq8.7\times10^{-15}$ cm$^{3}$ s$^{-1}$. Here, $n_{\rm col}$ is the effective collider density, 

Balancing CR pumping, radiative decay, and collisional transfer yields
\begin{align}
P_{10} + f_{12}  q_1 n_{\rm col} &= f_{10}  (A + q_2 n_{\rm col}), \\
P_{12} + f_{10}  q_2 n_{\rm col} &= f_{12}  (A + q_1 n_{\rm col}) .
\end{align}

Solving these coupled equations gives exact analytic expressions,
\begin{align}
f_{10} &=
\frac{(A + q_1 n_{\rm col}) P_{10} + q_1 n_{\rm col} P_{12}}
     {A  [A + (q_1 + q_2) n_{\rm col}]}, \\
f_{12} &=
\frac{(A + q_2 n_{\rm col}) P_{12} + q_2 n_{\rm col} P_{10}}
     {A  [A + (q_1 + q_2) n_{\rm col}]} ,
\end{align}
which reduce to simple, physically transparent limits in different density
regimes.

\noindent \textit{Critical densities:}
The densities at which collisional and radiative rates are comparable are $n_{c,1}=A/q_1\simeq3.9\times10^{6}$ cm$^{-3}$ and $n_{c,2}=A/q_2$, which depends strongly on temperature: $n_{c,2}\simeq10^{8}$ cm$^{-3}$ at 100 K, $\sim10^{10}$ cm$^{-3}$ at 50 K, and $\sim8\times10^{12}$ cm$^{-3}$ at 30 K.

\noindent\textit{Low-density regime ($n_{\rm col}\ll n_{c,1}$).} Collisions are negligible and radiative decay dominates, yielding $f_{10}\simeq P_{10}/A$ and $f_{12}\simeq P_{12}/A$. In this limit, the relative populations of ($v=1$, $J=2$) and ($v=1$, $J=0$) directly reflect the CR pumping ratio $P_{12}/P_{10}$, and no collisional redistribution occurs within the $v=1$ rotational manifold. Consequently, transitions originating from ($v=1$, $J=2$), such as 1--0 S(0) and 1--0 O(4), are not suppressed relative to 1--0 O(2).

\noindent\textit{High-density regime ($n_{\rm col}\gg n_{c,2}$).} Collisions dominate over radiative processes, giving $
\frac{f_{12}}{f_{10}} \simeq \frac{q_2}{q_1} = 5  \exp(-484 \text{ K}/T)$. At $T=100$ K this ratio is $\simeq0.04$, while at $T=30$ K it decreases to $\simeq5\times10^{-7}$, reflecting the strong temperature dependence of the upward collisional rate. The corresponding densities required to reach this limit are $n_{\rm col}\gtrsim A/q_2\simeq10^{8}$ cm$^{-3}$ at 100 K and $\gtrsim8\times10^{12}$ cm$^{-3}$ at 30 K.

\noindent\textit{Intermediate-density regime ($n_{c,1}\ll n_{\rm col}\ll n_{c,2}$).} 
This regime is likely the most relevant for Oph 163131. For illustrative purposes, we evaluate the scaling below using the T = 100 K collisional coefficients discussed above, while emphasizing that the relevant critical densities depend strongly on temperature. Here, de-excitation collisions efficiently drain ($v=1$, $J=2$) into ($v=1$, $J=0$), while radiative decay still controls the total $v=1$ population. The ratio simplifies to (for 100 K)
\begin{equation}
\frac{f_{12}}{f_{10}}
\simeq
\frac{P_{12}}{P_{10}}  
\frac{1}{1+n_{\rm col}q_1/A}
=
\frac{1.13}{1+n_{\rm col}/3.9\times10^{6}} .
\end{equation}

From Table \ref{tab:h2_line_fluxes}, the observed ratio 1--0 O(4)/1--0 O(2)$\simeq0.025$ is a factor $\sim8$ below the low-density CR prediction. Equating $1+n_{\rm col}/3.9\times10^{6}\simeq8$ gives $n_{\rm col}\sim3\times10^{7}$ cm$^{-3}$. This estimate should be regarded as illustrative rather than a formal density measurement; at lower temperatures $q_1$ decreases, shifting the inferred density upward, but accurate low-temperature rate coefficients are not yet available. Notably, the collider density inferred from this illustrative CR-based estimate is comparable to the densities at which the UV PDR models reproduce the observed 1--0 O(4)/1--0 O(2) and 1--0 S(0)/1--0 O(2) ratios (Figure \ref{fig:meudon_chi100_1000}), suggesting that collisional redistribution within the $v=1$ manifold operates in a similar density regime independent of the excitation mechanism.

In the intermediate- and high-density regimes, collisional redistribution within the $v{=}1$ manifold transfers population from $(v,J)=(1,2)$ into $(1,0)$. However, the resulting enhancement of the $(1,0)$ population is intrinsically limited.

In the low-density limit, the fractional populations are set purely by the non-thermal pumping rates, $f_{10}\simeq P_{10}/A$ and $f_{12}\simeq P_{12}/A$. In the opposite limit where $n_{\rm col}\gg n_{c,1}$, all population injected into $v{=}1$ rapidly cascades to $(1,0)$ prior to radiative decay, yielding $ f_{10}^{\rm max} \simeq \frac{P_{10}+P_{12}}{A}.$ The maximum enhancement of the $(1,0)$ population relative to the low-density case is therefore 
\begin{equation}
\frac{f_{10}^{\rm max}}{f_{10}^{\rm low}}
= 1+\frac{P_{12}}{P_{10}}.
\end{equation}

For CR secondary electrons, the ratio $P_{12}/P_{10}$ is of order unity ($\simeq1.1$ at $T\sim100$ K; e.g., \citealt{Gredel_infrared_H2_Xrays_1995}), implying a maximum enhancement factor $\lesssim2.1$. Thus, collisional redistribution can increase the 1--0 O(2) emissivity by at most a factor of $\sim2$.

This worked example illustrates how collisional redistribution within the $v=1$ manifold can suppress higher-$J$ para-H$_2$ emission under CR excitation while preserving vibrational excitation. The illustrative densities inferred here are broadly comparable to those required in the UV PDR models in Section \ref{sec:results:meudon_uv}, although the precise threshold depends sensitively on temperature and the uncertain low-temperature collider rates.


\end{document}